\documentclass[english,twocolumn,superscriptaddress,
  longbibliography]{revtex4-1}

\usepackage{soul}
\usepackage[dvipsnames]{xcolor}
\usepackage{float}
\usepackage{yhmath}
\usepackage{xcolor}
\usepackage{graphicx}
\usepackage{amsmath}
\usepackage{amsfonts}
\usepackage{color}
\usepackage{babel}
\usepackage{url}
\usepackage{hyperref}
\usepackage{booktabs}
\usepackage{tikz}
\usepackage{CJKutf8}
\usepackage[title]{appendix}
\usepackage{siunitx}

\definecolor{lime}{HTML}{A6CE39}
\DeclareRobustCommand{\orcidicon}{
	\begin{tikzpicture}
	\draw[lime, fill=lime] (0,0) 
	circle [radius=0.16] 
	node[white] {{\fontfamily{qag}\selectfont \tiny ID}};
	\draw[white, fill=white] (-0.0625,0.095) 
	circle [radius=0.007];
	\end{tikzpicture}
	\hspace{-2mm}
}

\foreach \x in {A, ..., Z}{\expandafter\xdef\csname orcid\x\endcsname{\noexpand\href{https://orcid.org/\csname orcidauthor\x\endcsname} {\noexpand\orcidicon}} }

\begin{document}

\title{Construction of Approximate Invariants Across Nonlinear Resonances}

\author{Yongjun Li\orcidA{}}\thanks{email: yli@bnl.gov}\affiliation{Brookhaven National Laboratory, Upton, New York 11973, USA}
\author{Derong Xu}\affiliation{Brookhaven National Laboratory, Upton, New York 11973, USA}
\author{Yue Hao}\affiliation{Michigan State University, East Lansing, Michigan  48864, USA}
\author{Sergei Nagaitsev}\affiliation{Brookhaven National Laboratory, Upton, New York 11973, USA}

\begin{abstract}
Our previous work [Y.~Li, D.~Xu, and Y.~Hao, Phys. Rev. Accel. Beams \textbf{28}, 074001 (2025)] introduced a square-matrix method for constructing approximate invariants recursively, order by order, from the one-turn map of non-integrable Hamiltonian systems. At fourth and higher even orders, the recursive construction is intrinsically non-unique because rotationally invariant polynomials generate a null space in the corresponding homological equations. In the resonant regime, this ambiguity can lead to an incorrect description of the resonance-island topology and is therefore particularly relevant to accelerator applications that rely on stable resonance islands. In this work, we select the null-space contributions by explicitly incorporating the null-space basis vectors and determining their coefficients through an averaging procedure over a prescribed region of interest. The method is
demonstrated using the Kobayashi map near a third-order resonance, extending the square-matrix construction to resonant Hamiltonian dynamics.
\end{abstract}

\maketitle

\section{Introduction}

In non-integrable Hamiltonian systems, such as those describing most circular accelerators, approximate invariants (hereafter,  invariants) provide a compact description of the phase-space topology and long-term nonlinear dynamics. Beyond their value as mathematical descriptors, such invariants have direct practical relevance to accelerator physics. In our previous work~\cite{li2025construction}, we proposed a square-matrix method for constructing invariants directly from the one-turn map. The method recursively determines the invariant order by order by solving the unit-eigenvalue problem of the transposed square matrix and was demonstrated for the National Synchrotron Light Source II (NSLS-II) storage ring.

However, the recursive construction is intrinsically non-unique at fourth and higher even orders. This issue was brought to our  attention by P.~Nishikawa~\cite{NishikawaPrivate}. At these orders, the coefficient matrix is singular because rotationally invariant polynomials span its null space, irrespective of whether the system is near a resonance. Consequently, the null-space coefficients cannot be determined from the homological equations alone, and infinitely many approximate invariants satisfy the same order-by-order construction. Away from resonance, this ambiguity may have little practical consequence for the phase-space topology of interest. In the resonant regime, however, the choice of null-space contribution becomes important because it can substantially alter the predicted resonance-island topology. Although a minimum-norm or pseudo-inverse solution selects a unique representative, it provides no systematic physical criterion for choosing the null-space contribution appropriate to the resonant dynamics. This limitation is particularly important in applications where the detailed structure and location of resonance islands play an essential role, including resonant  extraction~\cite{kobayashi1970theory, bryant2018resonant}, resonant trapping of scattered particles~\cite{cortes2026resonant}, and transverse resonance island beam manipulation~\cite{arion2018transverse, goslawski2016resonance, huschauer2017transverse, olsson2021studies, kim2022transverse, wang2023lattice}. In these applications, an accurate representation of the resonance-island topology is essential for describing and controlling the underlying nonlinear dynamics.

In this work, we determine the null-space contributions through an averaging procedure~\cite{zolkin2026geometry, Zolkin2026prab}. The optimal null-space coefficients are selected by minimizing the spatial variation of the leading-order residual of the approximate invariant over a prescribed region of interest (RoI). In this way, the resulting invariant can reproduce the resonance-island topology while preserving the recursive structure of the original square-matrix method. The approach is demonstrated using the Kobayashi map~\cite{kobayashi1970theory,dragt1982lectures} near a third-order resonance, where three stable resonance islands are present.

\section{Method}
  The evolution of nonlinear motion in a periodic Hamiltonian system, such as a circular accelerator, can be approximated by polynomial maps. Such maps can be extracted to arbitrary order using the truncated power series algorithm (TPSA)~\cite{berz1995modern,yang2009array,zhang2024cpptpsa}. As an illustrative example, we consider the Kobayashi map, which consists of a linear symplectic rotation followed by a thin nonlinear kick,
  \begin{align*}
    \begin{pmatrix}
    x_{n+1}\\
    p_{n+1}^{-}
    \end{pmatrix}
    &=
    \begin{pmatrix}
    \cos\mu+\alpha\sin\mu & \beta\sin\mu\\
    -\gamma\sin\mu & \cos\mu-\alpha\sin\mu
    \end{pmatrix}
    \begin{pmatrix}
    x_n\\
    p_n
    \end{pmatrix},\nonumber\\
    p_{n+1}
    &=p_{n+1}^{-}+b_4x_{n+1}^2+x_{n+1}^3,
  \end{align*}
  where the Twiss parameters satisfy $\beta\gamma-\alpha^2=1$. Although highly simplified, the Kobayashi map captures the essential features of nonlinear resonance dynamics and has been widely used as a benchmark for studying resonance crossing, stable islands, and invariants. 

\begin{figure}
        \centering
        \includegraphics[width=0.6\columnwidth]{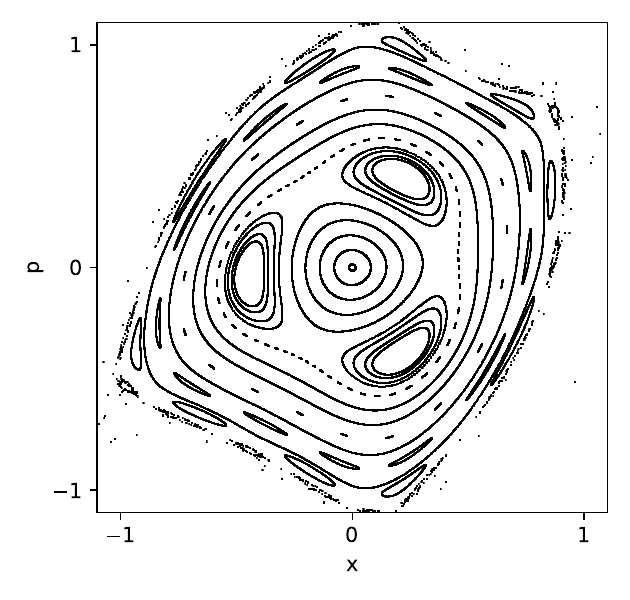}
        \caption{Phase-space portrait of the Kobayashi map near a third-order resonance.}
        \label{fig:Kobayashi_simu}
  \end{figure}

  With the following parameter set:
  \[
    \beta=\gamma=1,\quad \alpha=0,\quad
    \mu=2\pi\left(\frac{1}{3}+\frac{1}{100}\right),\quad
    b_4=\frac{1}{10},
  \]
  the phase-space trajectory is shown in Fig.~\ref{fig:Kobayashi_simu}. Even when the tune crosses the third-order resonance, the motion remains bounded over the simulated interval, cycling among the three stable resonance islands.

\subsection{Recursive construction}
  Following the method proposed in Ref.~\cite{li2025construction}, the one-turn map is represented by a truncated power series expansion,
    \begin{align}\label{eq:otm}
    x_{n+1} &= cx_n+sp_n, \nonumber\\
    p_{n+1} &= -sx_n+cp_n+\sum_{i+j=2}^{\Omega} h_{ij}x_n^ip_n^j,
    \end{align}
  where \(c=\cos\mu,\, s=\sin\mu,\) and $h_{ij}$ denotes the nonlinear coefficient of the monomial $x^ip^j$.

  Usually, a two-dimensional phase-space vector $\mathbf{z}=[x,p]^T$ is used to specify the particle coordinates at a given time/turn, where the superscript $^T$ denotes matrix or vector transpose. In our case, we first expand $\mathbf{z}$ to a monomial vector truncated $\mathbf{Z} $ at order $\Omega$,
    \begin{equation}\label{eq:Z}
    \mathbf{Z}=[x,p,x^2,xp,p^2,\ldots,p^{\Omega}]^T.
    \end{equation}
  The one-turn map can then be expressed in matrix form as
    \[
    \mathbf{Z}_{n+1}=\mathbf{M}\mathbf{Z}_n,
    \]
  where $\mathbf{M}$ is the square coefficient matrix derived from Eq.~\eqref{eq:otm}. The finite-dimensional matrix \(\mathbf{M}\) is not canonically symplectic in the usual sense. Nevertheless, the underlying polynomial map represented by \(\mathbf{M}\) preserves symplecticity through the retained polynomial order; any violation appears only in terms beyond the truncation order.
    
  An $\Omega^{th}$-order invariant is written as a polynomial
    \[
    W^{(\Omega)}=\mathbf{V}^T\mathbf{Z},
    \]
  where the superscript $^{(\Omega)}$ denotes the highest polynomial order retained, and $\mathbf{V}$ is the coefficient vector. Since $W^{(\Omega)}$ is invariant under the one-turn map, it satisfies
    \[
    \mathbf{V}^T\mathbf{Z}=\mathbf{V}^T\mathbf{M}\mathbf{Z},\qquad \forall\, \mathbf{Z},
    \]
  which immediately yields
    \[
    \mathbf{M}^T\mathbf{V}=\mathbf{V}.
    \]
  Therefore, the coefficient vector $\mathbf{V}$ is the eigenvector of $\mathbf{M}^T$ associated with the unit eigenvalue~\cite{chao2020lectures,forest2018beam}.

  Starting from the linear case, higher-order invariants are then constructed recursively. For the linear map, the monomial vector $\mathbf{Z}$ is truncated at second order, yielding a quadratic invariant $W^{(2)}$. Accordingly, the transposed coefficient matrix $\mathbf{M}^T$ is truncated to a $5\times5$ matrix. Solving the eigenvalue problem associated with the unit eigenvalue gives $\mathbf{V}^{(2)}=[0,0,1,0,1]^T$, i.e., the well-known Courant--Snyder invariant
  \[
  \label{eq:W2}
     W^{(2)}=\gamma x^2+2\alpha xp+\beta p^2=x^2+p^2.
  \]
      
  Next, cubic terms are added to \(W^{(2)}\) to form a third-order invariant. The corresponding coefficient vector is stacked as
    \[
    \mathbf{V}=\begin{bmatrix}
    \mathbf{V}^{(2)}\\
    \mathbf{V}^{(3)}
    \end{bmatrix},
    \]
  where $\mathbf{V}^{(3)}$ is the yet-to-be-determined coefficient vector. Since the upper-right block of $\mathbf{M}^T$ is identically zero, the matrix can be partitioned as
    \[
    \mathbf{M}^T=
    \begin{bmatrix}
    \mathbf{m}_0 & \mathbf{0}^{(3)}\\
    \mathbf{m}_2 & \mathbf{m}_3
    \end{bmatrix},
    \]
  where $\mathbf{m}_0$ is the quadratic block analyzed above, $\mathbf{0}^{(3)}$ is a zero matrix, and $\mathbf{m}_2$ and $\mathbf{m}_3$ are the lower-left and lower-right blocks, respectively. The subscripts of $\mathbf{m}_0$, $\mathbf{m}_2$, and $\mathbf{m}_3$
  label their positions in the block decomposition and do not denote the polynomial order. 

  The coefficient vector of the third-order invariant must satisfy
    \[
    \begin{bmatrix}
    \mathbf{V}^{(2)}\\
    \mathbf{V}^{(3)}
    \end{bmatrix}
    =
    \begin{bmatrix}
    \mathbf{m}_0 & \mathbf{0}^{(3)}\\
    \mathbf{m}_2 & \mathbf{m}_3
    \end{bmatrix}
    \begin{bmatrix}
    \mathbf{V}^{(2)}\\
    \mathbf{V}^{(3)}
    \end{bmatrix}.
    \]
  The first block equation, \(\mathbf{V}^{(2)}=\mathbf{m}_0\mathbf{V}^{(2)}\), is already satisfied by construction. Consequently, the second block equation can be solved independently, yielding
    \begin{equation}
    \mathbf{V}^{(3)}=(\mathbf{I}-\mathbf{m}_3)^{-1}\mathbf{m}_2\mathbf{V}^{(2)},
    \label{eq:iter}
    \end{equation}
  provided that $\mathbf{I}-\mathbf{m}_3$ is nonsingular.

  By solving Eq.~\eqref{eq:iter}, the cubic invariant is obtained as
    \[
    W^{(3)}=x^2+p^2-0.277877x^3-0.1x^2p+0.780012xp^2.
    \]
    
\subsection{Minimum-norm solution} 
  When proceeding to the fourth order, the situation changes. The homological equation for the invariant coefficients has the same form as Eq.~\eqref{eq:iter} 
    \begin{equation}
    \bigl(\mathbf{I}-\mathbf{m}_3\bigr)\mathbf{V}^{(4)}=\mathbf{m}_2\mathbf{V}^{(2,3)},
    \label{eq:fourth}
    \end{equation}
  where $\mathbf{V}^{(2,3)}=[\mathbf{V}^{(2)},\mathbf{V}^{(3)}]^T$,  $\mathbf{m}_3$ now is the $5\times 5$ block associated with the quartic monomials. However, the rotationally invariant polynomial 
  \begin{equation}\label{eq:w2}
    (x^2+p^2)^2=x^4+2x^2p^2+p^4,
  \end{equation}
  corresponds to the coefficient vector
  \[
    \mathbf{n}=[1,0,2,0,1]^T,
  \]
  which spans the null space of \(\mathbf{I}-\mathbf{m}_3\). As a result, the fourth-order invariant is determined only up to an arbitrary multiple of \((x^2+p^2)^2\), and an additional normalization condition is required.

  To obtain a particular solution of the fourth-order homological equation, we first compute the minimum-norm~\cite{golub1996matrix} solution \(W^{(4,0)}\). This is accomplished by removing one linearly dependent row from Eq.~\eqref{eq:fourth} and then augmenting the system with an orthogonality constraint,
    \begin{equation*}
    \begin{bmatrix}
    \widetilde{\mathbf I-\mathbf m_3}\\
    \mathbf n^{T}
    \end{bmatrix}
    \mathbf V^{(4)}
    =
    \begin{bmatrix}
    \widetilde{\mathbf m_2\mathbf V^{(2,3)}}\\
    0
    \end{bmatrix}.
    \end{equation*}
  where $\widetilde{\mathbf I-\mathbf m_3}$ and $\widetilde{\mathbf m_2\mathbf V^{(2,3)}}$ denote the coefficient matrix and right-hand-side vector after removing one linearly dependent equation, and \(\mathbf{n}\) spans the null space of \(\mathbf{I}-\mathbf{m}_3\). The additional equation
    \[
    \mathbf{n}^T\mathbf{V}^{(4)}=0,
    \]
  forces \(\mathbf{V}^{(4)}\) to be orthogonal to the null space, thereby eliminating the arbitrary homogeneous component and yielding the unique solution orthogonal to the null space and hence the minimum-norm solution of the consistent homological equation. The general fourth-order invariant, accounting for the nontrivial null space of \(\mathbf{I}-\mathbf{m}_3\), is
    \begin{equation}\label{eq:f4}
    W^{(4)}=W^{(4,0)}+c_4(x^2+p^2)^2,
    \end{equation}
  where \(c_4\) cannot be determined from the fourth-order homological equation alone. It is worthwhile to note that, if the quadratic invariant contains a nonzero cross term (i.e., \(\alpha \neq 0\)), the form of the null vector \(\mathbf{n}\) must be modified as \([\gamma^2,4\alpha\gamma,2\beta\gamma+4\alpha^2,4\alpha\beta,\beta^2]\) to remain consistent with the corresponding quadratic invariant.

\subsection{Approximate invariants near resonance}
\label{sec:resonant_invariant}

The null-space ambiguity described above is algebraically generic and is present irrespective of resonance. Its physical consequence, however, becomes particularly important near a low-order resonance, where different representatives of a truncated approximate invariant can lead to substantially different descriptions of the resonance-island topology. To clarify this point, consider a one-degree-of-freedom (1-DoF) map truncated at polynomial order $\Omega$. Away from resonances of order $\Omega$ or lower, a formal invariant may be written as
\[
W_1=W_{\rm CS}+\Delta W,
\label{eq:formal_W1}
\]
where $W_{\rm CS}$ is the Courant--Snyder invariant and $\Delta W$ contains nonlinear corrections. Other formal invariants generated from the same fundamental invariant are simply functions of $W_1$, for example,
\[
W_n=W_1^n,\qquad n \leq \left\lfloor \frac{\Omega}{2} \right\rfloor,
\label{eq:formal_Wn}
\]
where $\lfloor \Omega/2 \rfloor$ denotes the largest integer not exceeding $\Omega/2$. Thus, these quantities do not constitute additional functionally independent invariants; rather, they represent different parameterizations of the same family of invariant level sets. In the square-matrix representation, this functional freedom appears directly as a null space of the homological operator at even polynomial orders. The repeated even-order singularities can therefore be understood as the finite-order algebraic counterpart of the functional freedom represented by $W_1^2,\,W_1^3,\,\dots$ in the truncated Koopman space~\cite{Koopman1931}. 

Near an $m^{th}$-order resonance, let the phase advance satisfy
\[
m\mu=2\pi(k+\epsilon),\qquad |\epsilon|\ll 1,
\]
where $k$ is an integer. In this limit, the perturbative representation of the conventional one-turn formal invariant \(W_1\) generally develops small-denominator terms and becomes ill conditioned as $\epsilon\rightarrow0$. A regular description of the resonant dynamics is
instead obtained from the corresponding $m$-turn map~\cite{dragt1982lectures}. Equivalently, taking the logarithm of the Koopman representation of the $m$-turn map yields a pseudo-Hamiltonian $H$ that can be expressed as a nonlinear function of the one-turn invariant,
\begin{equation}
H=\sum_{n=1}^{\left\lfloor\Omega/2\right\rfloor}
a_n W_1^n 
\label{eq:res_pseudoH}
\end{equation}
where the coefficients $a_n$ are fixed by the resonant normal-form construction. The pseudo-Hamiltonian $H$ generates the slow resonant dynamics of the $m$-turn map and remains well behaved when the one-turn representation becomes ill conditioned. Equation~\eqref{eq:res_pseudoH} is introduced here only to illustrate the relation between the one-turn formal invariant and the resonant pseudo-Hamiltonian; in the present work, $H$ is constructed directly from the $m$-turn Koopman operator, as described in Appendix~\ref{app:koopman}.

It is useful to distinguish the construction in Eq.~\eqref{eq:res_pseudoH} from the intrinsic parameterization freedom of the one-turn invariant. Since any function of an exact invariant is itself an invariant, the transformation
\begin{equation}
W_1\longrightarrow \widetilde W_1
=\sum_{n=1} d_n W_1^n,
\label{eq:invariant_reparam}
\end{equation}
does not produce a functionally independent exact invariant. Provided that the transformation is locally monotonic, $W_1$ and $\widetilde W_1$ describe exactly the same invariant level sets. The coefficients $d_n$ in Eq.~\eqref{eq:invariant_reparam} therefore characterize a reparameterization freedom of the invariant, whereas the coefficients $a_n$ in Eq.~\eqref{eq:res_pseudoH} select a particular function of $W_1$ that serves as the pseudo-Hamiltonian generating the $m$-turn dynamics.

For a truncated non-integrable map, however, this formal equivalence is no longer exact. Truncation of different parameterizations generally produces different approximate invariants. Algebraically, this freedom appears through null-space terms such as $c_4(x^2+p^2)^2$ in Eq.~\eqref{eq:f4}. Away from resonance, different choices of these coefficients may have little visible effect in the central Kolmogorov--Arnold--Moser (KAM) region. Near a low-order resonance, however, truncation errors can be strongly amplified, and different representatives may give substantially different approximations to the resonance-island structure. The null-space coefficients therefore provide useful degrees of freedom for optimizing the approximate invariant over the phase-space RoI, rather than being fixed solely by an arbitrary algebraic normalization.

\subsection{Residual averaging procedure}

Finally, we adopt an averaging procedure, motivated by Refs.~\cite{zolkin2026geometry,Zolkin2026prab}, to determine the null-space coefficient $c_4$. Substituting the one-turn map, Eq.~\eqref{eq:otm}, into the quartic approximate invariant, Eq.~\eqref{eq:f4}, gives the leading fifth-order residual
\begin{equation}
\label{eq:f5}
f_5(x,p;c_4)
=
\sum_{i+j=5}
\left(a_{ij}+c_4 b_{ij}\right)x^i p^j,
\end{equation}
where the coefficients $a_{ij}$ and $b_{ij}$ are listed in Tab.~\ref{tab:f5_coeff}. The polynomial $f_5$, as the leading one-turn invariance defect of the quartic candidate invariant \(W^{(4)}\), quantifies the first non-vanishing local departure from conservation under the map.

\begin{table}[H]
\centering
\caption{Coefficients of the fifth-order residual polynomial in
Eq.~\eqref{eq:f5}.}
\label{tab:f5_coeff}
\setlength{\tabcolsep}{4pt}
\begin{tabular}{crr}
\hline
Monomial & $b_{ij}$ & $a_{ij}$ \\
\hline
$x^5$      & $-0.102030$ & $-0.100855$ \\
$x^4p$     & $ 0.239347$ & $ 0.641574$ \\
$x^3p^2$   & $-0.129108$ & $-1.247891$ \\
$x^2p^3$   & $ 0.085779$ & $ 0.591625$ \\
$xp^4$     & $-0.027078$ & $ 0.537605$ \\
$p^5$      & $-0.153568$ & $-0.408030$ \\
\hline
\end{tabular}
\end{table}

The averaging is performed over a prescribed RoI,
\[
\mathcal{R}
=
[x_{\min},x_{\max}]
\times
[p_{\min},p_{\max}],
\]
which specifies the portion of phase space over which the approximate invariant is required to provide the most faithful description of the dynamics. The choice of RoI therefore provides a physical weighting for selecting among the otherwise undetermined null-space contributions. For the present map, we choose
\[
\mathcal{R}=[-1,1]\times[-1,1],
\]
which contains the central KAM region and the primary resonance islands shown in Fig.~\ref{fig:Kobayashi_simu}.

\begin{figure}
\centering
\includegraphics[width=0.43\columnwidth]{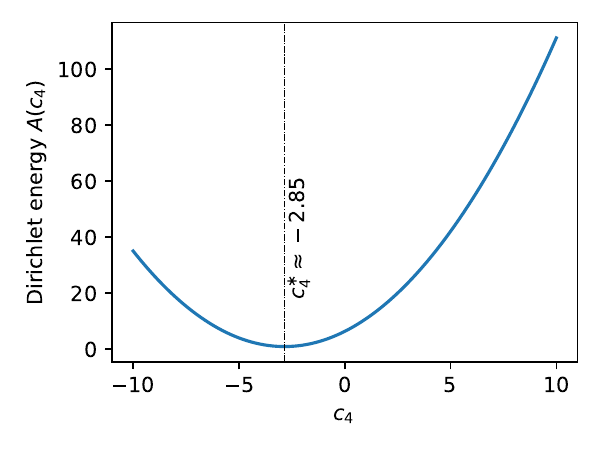}
\includegraphics[width=0.55\columnwidth]{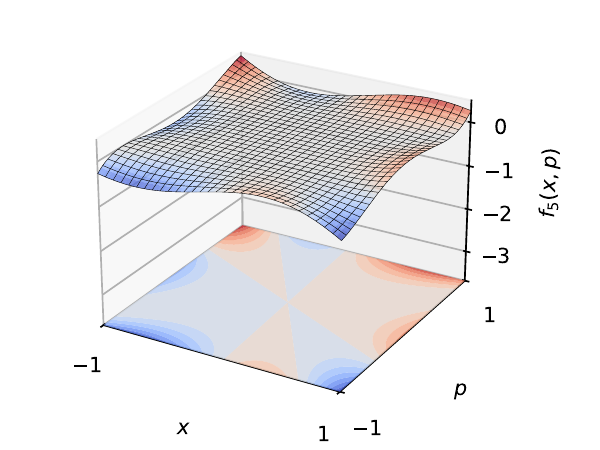}
\caption{Determination of the optimal quartic coefficient
$c_4^\ast$. Left: Dirichlet energy $A(c_4)$ of the leading
fifth-order residual over the RoI, $[-1,1]\times[-1,1]$, as a
function of $c_4$. The minimum determines $c_4^\ast$. Right:
the corresponding residual surface $f_5(x,p;c_4^\ast)$.}
\label{fig:f5_c4}
\end{figure}

Geometrically, \(f_5\) determines the leading displacement of the level sets of $W^{(4)}$ after one turn. Its gradient measures the spatial non-uniformity of that displacement over the RoI. Since it is a homogeneous fifth-order polynomial and therefore vanishes at the origin, its magnitude at a single point does not provide a suitable optimization criterion. To determine $c_4$, we penalize rapid spatial variation of $f_5$ over the RoI using the Dirichlet energy~\cite{Evans2010PDE},
\begin{equation*}
\label{eq:dirichlet}
A(c_4)
=
\int_{x_{\min}}^{x_{\max}}
\int_{p_{\min}}^{p_{\max}}
\left|\nabla f_5(x,p;c_4)\right|^2
\,\mathrm{d}p\,\mathrm{d}x,
\end{equation*}
where
\[
|\nabla f_5|^2
=
\left(\frac{\partial f_5}{\partial x}\right)^2
+
\left(\frac{\partial f_5}{\partial p}\right)^2.
\]
The optimal null-space coefficient is therefore defined by
\[
c_4^\ast
=
\arg\min_{c_4} A(c_4).
\]
Because $f_5$ depends linearly on $c_4$, $A(c_4)$ is quadratic
in $c_4$ and has a unique minimum. For the RoI specified above,
we obtain
\[
c_4^\ast\approx-2.850159.
\]
Fig.~\ref{fig:f5_c4} shows the Dirichlet energy as a function
of $c_4$ and the corresponding residual surface at the minimum.

Because the averaging criterion is defined over a finite phase-space region, the optimized coefficient generally depends on the choice of RoI, which specifies where the spatial variation of the residual is minimized. Nevertheless, the result should not be overly sensitive to reasonable variations of the RoI that continue to encompass the stable dynamics of
interest. For the present example, numerical tests show that $c_4^\ast$ remains nearly unchanged when the RoI is uniformly rescaled from $[-0.5,0.5]^2$ to $[-2,2]^2$, as expected from the homogeneity of the leading residual. Similar robustness is observed for nonuniform and shifted RoIs, provided that they encompass the central KAM region and the primary resonance islands. These results suggest that the selected null-space contribution is governed primarily by the local resonant dynamics rather than by the precise boundaries of the averaging region.

The same procedure applies recursively at all subsequent even orders. At orders $6,8,\ldots$, the homological equations again contain null-space directions associated with the rotationally
invariant polynomials
\[
(x^2+p^2)^3,\qquad
(x^2+p^2)^4,\qquad \ldots ,
\]
whereas the corresponding odd-order systems remain non-singular because no rotationally invariant polynomial of odd degree exists. Using the same RoI and averaging criterion gives
\[
c_6^\ast \approx 0.254665,
\qquad
c_8^\ast \approx -1.911989,
\qquad \ldots .
\]

\begin{figure}
\centering
\includegraphics[width=0.95\columnwidth]{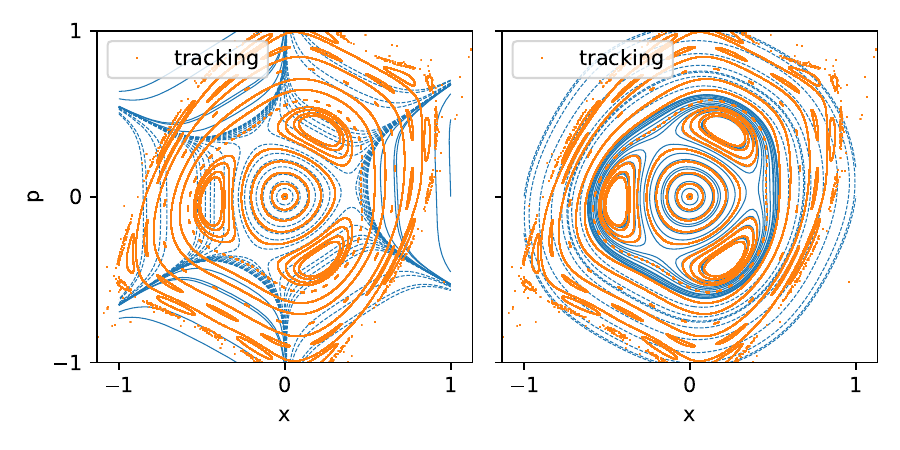}
\caption{Comparison of the eighth-order invariants with tracking simulations. The left panel shows the invariant constructed without the null-space corrections, while the right panel includes the null-space coefficients determined by the averaging procedure. Blue dashed curves denote invariant contours, and orange dots denote tracking results.}
\label{fig:Kobayashi}
\end{figure}

The resulting eighth-order invariant is compared with tracking simulations in the right panel of Fig.~\ref{fig:Kobayashi}. With the optimized null-space coefficients included, the invariant reproduces the third-order resonance-island topology and agrees well with the tracking results. For comparison, setting
\[
c_{4,6,8,\ldots}^{\ast}=0
\]
selects the minimum-norm representatives without the null-space corrections. As illustrated in the left panel, although this choice also reproduces the central KAM region, it fails to capture the resonance-island topology. This comparison illustrates why the null-space ambiguity, while often inconsequential away from resonance, becomes important when the detailed resonant topology is of interest. This also explains why the ambiguity was not apparent in our previous application to the NSLS-II storage ring~\cite{li2025construction}: its tune is sufficiently far from low-order resonances, and the central KAM region is already well reproduced without explicitly accounting for the null-space contributions.

The coefficients of the resulting square-matrix invariant through sixth order are compared in Tab.~\ref{tab:appinv8} with those of the Koopman-based pseudo-Hamiltonian described in Appendix~A. The two constructions are normalized to the same quadratic Courant-Snyder invariant and agree exactly through cubic order. Their higher-order coefficients are also close, while the fourth-order difference satisfies the \(1:2:1\) ratio associated with \((x^2+p^2)^2\). Thus, the remaining difference is consistent with the intrinsic null-space freedom of the recursive construction rather than a change in the underlying resonance-island topology. The corresponding invariant contours differ only weakly over the RoI.

\begin{table}[htb]
\centering
\caption{Comparison of the coefficients obtained from the
Koopman-based pseudo-Hamiltonian and the square-matrix
approximate invariant through sixth order.}
\label{tab:appinv8}
\begin{tabular}{crr}
\hline
Monomial & Koopman-operator & Square-matrix \\
\hline
$x^2$      & $ 1.000000$ & $ 1.000000$ \\
$p^2$      & $ 1.000000$ & $ 1.000000$ \\

$x^3$      & $-0.277877$ & $-0.277877$ \\
$x^2p$     & $-0.100000$ & $-0.100000$ \\
$xp^2$     & $ 0.780012$ & $ 0.780012$ \\

$x^4$      & $-2.868827$ & $-2.822153$ \\
$x^3p$     & $-1.078001$ & $-1.078001$ \\
$x^2p^2$   & $-5.642616$ & $-5.549268$ \\
$p^4$      & $-3.226939$ & $-3.180264$ \\

$x^5$      & $ 0.071147$ & $ 0.045208$ \\
$x^4p$     & $-0.215751$ & $-0.225085$ \\
$x^3p^2$   & $-0.375778$ & $-0.328904$ \\
$x^2p^3$   & $ 0.645388$ & $ 0.636053$ \\
$xp^4$     & $ 0.250411$ & $ 0.323224$ \\

$x^6$      & $-0.080458$ & $-0.129950$ \\
$x^5p$     & $ 5.680194$ & $ 5.582158$ \\
$x^4p^2$   & $ 1.481695$ & $ 1.311512$ \\
$x^3p^3$   & $ 6.403795$ & $ 6.295884$ \\
$x^2p^4$   & $-0.097379$ & $-0.252827$ \\
$p^6$      & $ 2.133719$ & $ 2.047194$ \\
$\vdots$   & $\vdots$   & $\vdots$ \\
\hline
\end{tabular}
\end{table}

\section{Discussion}
  The integration of the averaging procedure into the recursive square-matrix method provides a systematic criterion for selecting among infinitely many approximate invariants. Several aspects of the proposed method merit further discussion, including the non-uniqueness of approximate invariants, its extension to multi-degree-of-freedom systems, and its applicability to accelerator lattice design.

  \textbf{Non-uniqueness of approximate invariants:} Constructed invariants are not unique in two respects. First, they can be obtained using different approaches. For the 1-DoF system considered here, for example, a pseudo-Hamiltonian derived from the Koopman operator approach (Appendix~\ref{app:koopman}) also serves as an approximate invariant. Despite their different constructions, the two approaches yield generally similar, but not identical, coefficients, as shown in Table~\ref{tab:appinv8}. Second, approximate invariants are not uniquely defined even within a given construction. In the present method, adopting alternative measures of the residual or different RoIs could therefore lead to different approximate invariants.
  
  \textbf{Extension to multi-degree-of-freedom systems:} The numerical demonstrations in this work are intentionally restricted to one degree of freedom, where the null-space ambiguity and its resolution through averaging can be isolated without the additional complications associated with coupled resonances and high-dimensional phase-space topology. The algebraic construction itself extends formally to systems with multiple degrees of freedom. However, its application to coupled resonant dynamics involves additional considerations and is beyond the scope of the present work. For example, for two-degree-of-freedom transverse motion in a circular accelerator, the null space associated with the quartic-order equation is three-dimensional and is spanned by
  \[
  W_x^2,\qquad W_xW_y,\qquad W_y^2,
  \]
  where
  \[
  W_u=\gamma_u u^2+2\alpha_u u p_u+\beta_u p_u^2,
  \qquad u\in{x,y}.
  \]
  The corresponding null-space coefficients can be determined by extending the same averaging criterion from the optimization of a single null-space coefficient to a joint optimization over the multidimensional null space
  \[
  c_{xx}W_x^2+c_{xy} W_xW_y+c_{yy}W_y^2.
  \]
  A detailed investigation of this extension, including the effects of linear and nonlinear coupling resonances, is left for future work.

  \textbf{Applications to accelerator lattice design:} The even-order singularity and the associated null-space ambiguity are present irrespective of resonance. For most circular accelerators operated away from low-order resonances, however, this ambiguity has limited practical impact on the central KAM region, which is typically the phase-space structure of primary interest. In practice, machine tunes are deliberately chosen to avoid low-order resonances and maximize this central stable region. Stable resonance islands are instead exploited in specialized applications, including resonant extraction and resonance-island-based synchrotron-radiation techniques. These applications are often dominated by a single-plane resonance, for which a 1-DoF reduction provides the leading description of the resonant dynamics and can be treated using the method developed here. Some special cases, however, require extensions of the present approach. One example arises when the linear tune lies exactly on a resonance. In that case, the null space of the square sub-block \(\mathbf{I}-\mathbf{m}_3\) can acquire additional dimensions, requiring additional null-space basis vectors in the construction of the approximate invariant. Another example is the integrable McMillan-like map with a saddle fixed point at the origin~\cite{nagaitsev2020}, for which the Courant--Snyder invariant is not well defined at small amplitudes. These cases, however, lie beyond the scope of conventional accelerator lattice design and optimization.
  
\appendix

\section{Construction of a pseudo-Hamiltonian using the Koopman operator}\label{app:koopman}

Although independent of the invariant construction presented in the main text, the Koopman-operator~\cite{Koopman1931} approach provides an alternative route to a pseudo-Hamiltonian, which in a 1-DoF system also serves as an approximate invariant. In particular, it makes the non-uniqueness of approximate invariants explicit. Near resonance, however, evaluating the matrix logarithm requires a branch-selection procedure, as described below.

Let $\mathbf{M}$ denote the square-matrix representation of the one-turn map acting on the monomial basis
\[
\mathbf{Z}
=
[1,x,p,x^2,xp,p^2,\cdots]^T,
\]
Here, the first component $1$ represents the constant basis function. It is omitted in Eq.~\eqref{eq:Z} in the main text because the reference orbit is chosen as the origin. For a polynomial function
\[
W(x,p)=\mathbf{V}^T\mathbf{Z},
\]
the one-turn transformation is
\[
\mathbf{V}^T\mathbf{M}\mathbf{Z}
=
(\mathbf{M}^T\mathbf{V})^T\mathbf{Z},
\]
showing that $\mathbf{M}^T$ is the matrix representation of the Koopman operator acting on the coefficient vector.

If the one-turn map admits a Lie representation~\cite{dragt1982lectures},
\[
\mathcal{M}
=
\exp(-:H:),
\]
where $H(x,p)$ is the pseudo-Hamiltonian, then $\mathbf{M}^T$ and the Lie operator induce the same transformation on polynomial functions. Consequently~\cite{forest2016tracking},
\[
\mathbf{L}=\log(\mathbf{M}^T),
\]
corresponds to the Lie operator
\[
\mathbf{L} \Longleftrightarrow -:H: .
\]

To reconstruct the pseudo-Hamiltonian, we first apply $\mathbf{L}$ to the coefficient vectors corresponding to the coordinate functions,
\[
\mathbf{e}_x=[0,1,0,\cdots]^T,\qquad
\mathbf{e}_p=[0,0,1,\cdots]^T,
\]
where unlisted elements are zeros. Defining
\[
\mathbf{g}_x=\mathbf{L}\mathbf{e}_x,
\qquad
\mathbf{g}_p=\mathbf{L}\mathbf{e}_p,
\]
the corresponding polynomial functions are
\[
G_x=\mathbf{g}_x^T\mathbf{Z},
\qquad
G_p=\mathbf{g}_p^T\mathbf{Z}.
\]
Since
\[
-:H:x=\frac{\partial H}{\partial p},
\qquad
-:H:p=-\frac{\partial H}{\partial x},
\]
we obtain
\begin{equation}\label{eq:Hamiltonian}
\frac{\partial H}{\partial p}=G_x,
\qquad
\frac{\partial H}{\partial x}=-G_p.
\end{equation}
Therefore,
\[
dH=-G_p\,dx+G_x\,dp.
\]
For an exact symplectic representation, the one-form above is closed. Within the finite-order truncation adopted here, any non-closure is of higher order than the retained terms. The pseudo-Hamiltonian can therefore be reconstructed by path integration,
\[
H(x,p)
=
\int_{\gamma}
\left(
-G_p\,dx
+
G_x\,dp
\right),
\]
where $\gamma$ is any path connecting a reference point to $(x,p)$.

Near a low-order resonance, direct evaluation of $\log(\mathbf{M}^T)$ suffers from the multi-valued nature of the complex logarithm. For the third-order resonance considered in this work, this ambiguity can be removed by evaluating the logarithm of the three-turn map~\cite{dragt1982lectures},
\[
\mathbf{N}
=
\left(\mathbf{M}^T\right)^3,
\]
whose eigenvalues are sufficiently close to unity to permit a unique branch selection. The pseudo-Hamiltonian is then obtained from
\[
\mathbf{L}
=-\frac{1}{3\mu-2\pi}\log(\mathbf{N})=
-\frac{100}{6\pi}
\log(\mathbf{N}),
\]
followed by the reconstruction procedure described above. The resulting pseudo-Hamiltonian coefficients are listed in Tab.~\ref{tab:appinv8} for comparison with the invariant obtained using the averaging procedure.

Interestingly, the difference between the two constructions at fourth order still follows the ratio $1:2:1$:
\[
\Delta C_{x^4}:\Delta C_{x^2p^2}:\Delta C_{p^4}= 0.046674:0.093348:0.046674,
\]
and thus lies precisely within the intrinsic null-space freedom of the recursive construction identified in Eq.~\eqref{eq:w2}. This suggests that the Koopman-operator construction implicitly selects the null-space contribution slightly differently from the averaging procedure adopted here.


\section{Invariant contours versus Hamiltonian flow}

The approximate invariant constructed in the main text and the pseudo-Hamiltonian obtained from the Koopman operator in Appendix~\ref{app:koopman} provide scalar functions whose level sets reproduce the phase-space topology of the nonlinear map. However, this geometrical property should be distinguished from the dynamical role of a true Hamiltonian. An invariant of a discrete symplectic map constrains the map to approximately preserve its level sets, but does not uniquely determine how points are transported along or between them. Therefore, the one-turn dynamics cannot in general be recovered by treating the invariant as a Hamiltonian and integrating Eq.~\eqref{eq:Hamiltonian}.

\begin{figure}[t]
 \centering
 \includegraphics[width=\columnwidth]{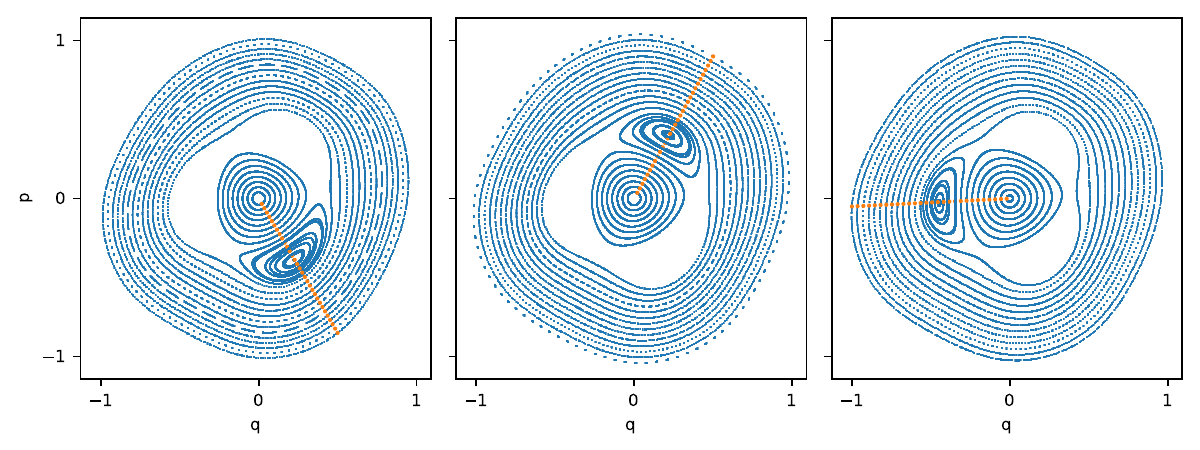}
 \caption{Trajectories generated by the continuous Hamiltonian flow associated with the constructed invariant. Blue trajectories are generated by integrating Hamilton's Eqs.~\eqref{eq:Hamiltonian} from the initial conditions indicated by the yellow dots. Within the resonance-island region, trajectories remain trapped in a single island, in contrast to the cyclic motion among islands under the original one-turn map in Fig.~\ref{fig:Kobayashi_simu}.}
 \label{fig:single_island_flow}
\end{figure}

This distinction is particularly clear near this stable third-order resonance. The one-turn map $M$ cyclically permutes the three resonance islands, whereas the $M^3$ map maps each island onto itself. However, the invariants extracted in the main text and Appendix~\ref{app:koopman} don't contain the discrete island-to-island permutation if interpreted as Hamiltonian flow. Fig.~\ref{fig:single_island_flow} illustrates this distinction visually. The blue trajectories are obtained by treating the invariant as a continuous Hamiltonian and integrating Eq.~\eqref{eq:Hamiltonian}. These trajectories within the resonance-islands remain confined to individual islands, although the contour of invariants reproduces the resonance-island topology.

This limits the direct application of the invariant-flow method developed in Ref.~\cite{Xu2026} for extracting fundamental frequencies. To extract the oscillation frequency around stable resonance fixed points that are cyclically permuted by the one-turn map, one can introduce a co-moving frame that factors out the discrete island-to-island permutation. Such a construction, however, would require prior knowledge of the resonance structure, including its order and the corresponding periodicity. For example, third- and fourth-order resonances naturally involve three- and four-turn periodicities, respectively. Thus, the appropriate co-moving representation does not follow automatically from the one-turn map without first identifying the underlying resonance. Whether such a representation can provide a systematic connection between the invariant flow and the original one-turn dynamics remains an open question.

\begin{acknowledgments}
 The authors thank Patrice Nishikawa (formerly of KEK) for valuable discussions, insightful comments, and continued support of this work. In particular, he introduced us to the Koopman-operator-based construction of the pseudo-Hamiltonian described in Appendix~\ref{app:koopman}, carried out the corresponding invariant calculation used for comparison with the square-matrix method, and carefully proofread the manuscript. His contributions also helped clarify several aspects of the formulation and interpretation of formal and approximate invariants in the vicinity of nonlinear resonances, as discussed in Sec.~\ref{sec:resonant_invariant}. The authors also thank Hao-Wen Luo (NSRRC) and Zhe Duan (IHEP) for fruitful discussions. This work was supported by Brookhaven Science Associates, LLC, under Contract No.~DE-SC0012704 with the U.S. Department of Energy (DOE); by the DOE Office of Basic Energy Sciences through Field Work Proposal 2025-BNL-PS040 (Y.L.); by a DOE Early Career Research Program award (D.X.); and by the DOE Office of High Energy Physics under Award No.~DE-SC0019403 (Y.H.).
\end{acknowledgments}

\bibliography{ref.bib}

\end{document}